# Solar Sail: Materials and Space Environmental Effects[1]

**Roman Ya. Kezerashvili**

*New York City College of Technology, The City University of New York, 300 Jay Street, Brooklyn, NY 11201, USA*
*The Graduate School and University Center, The City University of New York, 365 Fifth Avenue, New York, NY 10016, USA*

Theoretical aspects of a solar sail material degradation are presented when the solar electromagnetic and corpuscular forms of radiation were considered as sources of degradation. The analysis of the interaction of two components of solar radiation, the electromagnetic radiation and radiation of low- and high-energy electrons, protons, and helium ions emitted by the Sun with the solar-sail materials is discussed. The physical processes of the interactions of photons, electrons, protons and α-particles with sail material atoms and nuclei, leading to the degradation and ionization of solar sail materials are analyzed. The dependence of reflectivity and absorption for solar sail materials on temperature and on wavelength of the electromagnetic spectrum of solar radiation is investigated. It is shown that the temperature of a solar sail increases approximately as $T \sim r^{-2/5}$, with the decrease of the heliocentric distance $r$, when are taking into account the temperature dependence of optical parameters of the solar sail material.

## Nomenclature

| | | |
|---|---|---|
| $\rho$ | = | density of the sail material, g/cm$^3$ |
| $d$ | = | thickness of solar sail material |
| $\sigma(\omega,T)$ | = | electrical conductivity, $\Omega^{-1}\text{m}^{-1}$ |
| $\sigma_0$ | = | electrical DC conductivity at temperature $T_0$, $\Omega^{-1}\text{m}^{-1}$ |
| $\rho(\lambda,T)$ | = | spectral reflectivity |
| $\alpha(\lambda,T)$ | = | spectral absorptivity |
| $\tau(\lambda,T)$ | = | spectral transmissivity |
| $\varsigma(\lambda,T)$ | = | spectral emissivity |
| $\varsigma(T)$ | = | emissivity |
| $\sigma_{SB}$ | = | Stefan-Boltzman constant, Wm$^2$K$^{-4}$ |
| $\mu$ | = | mass attenuation coefficient |
| $n$ | = | number of metal atoms in the unit volume, cm$^{-3}$ |
| $Z$ | = | atomic number |
| $T$ | = | temperature, K |
| $\varepsilon(\omega)$ | = | dielectric function |

## I. Introduction

The solar sail is one of the very few proposed space-propulsion methods that has enormous potential, because it takes advantage of sunlight and does not require the chemical fuel that spacecraft currently relies on for propulsion. Solar sails accelerate slowly but surely, capable of eventually reaching tremendous speeds that may eventually be applied to interstellar exploration and travel. In fact, many scientists consider solar sailing the only reasonable way to make interstellar travel a reality. Usually a solar sail is considered as a thin membrane that uses the momentum carried by electromagnetic radiation originated from the sun to propel a spacecraft. However, the Sun space environment is a very dynamic place. As well as a high photon flux, there is a stream of electrically charged particles that is ejected from the Sun. This corpuscular part of the solar radiation is highly variable in terms of both velocity and density. The interactions of a solar-sail material with positive and negative Sun-generated particles must therefore be considered by interstellar mission planners because a solar sail will be a long term under an influence of different type space environmental effects. Solar energetic particle events can have a significant effect on both the operations and design of a solar-sail spacecraft. This was one of the reasons for formation of the Environmental Effects Group at NASA's Marshall Space Flight Center [1] that is tasked with characterizing the material properties of newly developed sail materials and further characterizing these materials in emulated space environments. Furthermore, the authors of Ref. [2] established in November 2004 the Solar Sail Degradation Model Working Group to take a step forward towards to study the general optical degradation behavior of solar sails and its impact on mission analysis. However, it is interesting to mention that more than four decades ago NASA experience has indicated a need for uniform criteria for the design of space vehicles related to effect of space environment on its material [3]. Present-day investigations [4-6] indicated that the space environmental effects degrade sail material thermo-optical properties and the sail material mechanical stability with increasing radiation exposure. Solar Sail Materials [7] is an on-going project for the European Space Agency relying on past and recent European solar sail design projects. It aims at developing and testing future technologies suitable for large, operational solar sailcrafts.

---

[1] Talk given at 3$^{rd}$ International Symposium on Solar Sailing, ISSS 2013, June 11-13, 2013, Glasgow, UK

A solar-sail performance is significantly affected by four factors: the areal mass of the sail, the optical properties of the sail film, the mechanical properties of the sail films, and the sail geometry. The first three factors are depending on each other that makes it difficult to study their influence on the performance of solar sail. The optical parameters such as a radiation absorption coefficient and emissivity are the major parameters governing the solar sail's surface temperature and its serviceability. This article focuses on theoretical aspects of a solar sail material degradation when the solar electromagnetic and corpuscular forms of radiation were considered as sources of degradation. The analysis of the interaction of two components of solar radiation, the electromagnetic radiation and radiation of low- and high-energy electrons, protons, and helium ions emitted by the Sun with the solar-sail material that lead to it degradation is addressed.

The remainder of this paper is organized in the following way. In Sec. II we discuss the structure and energy spectrum of the solar radiations. The temperature dependence of a solar sail parameters is discussed in Sec. III, where is presented the minimal thickness of the solar sail film that provides the maximum reflection of the solar radiation, as well as is given the dependence of the temperature of the solar sail material on the heliocentric distance when the optical parameters depend on temperature. The results of interaction of the ultraviolet radiation with the solar sail material are presented in Sec. IV A, while Sec. IV B has deal with the degradation of the solar sail material by solar electrons, protons and ions of helium. Finally, in Sec. V we summarize our studies and present the conclusions.

## II. Solar Radiation

Solar radiation has two components: the electromagnetic radiation and the corpuscular radiation that consist from low- and high-energy elementary particles like electron, protons, neutrinos and ions of light nuclei emitted by the Sun. The solar corpuscular radiation mainly results from a solar wind, coronal mass ejections, solar flares, and solar prominences that are the main sources of electrons and protons. The first two produce electrons with energies from about 0.01 eV up to a few hundreds of eV and the flares are a source of high energy electrons with energies from 1 MeV up to hundreds of MeV. The energy spectrum of the solar protons extends from 0.2 keV to a few tens of keV for solar wind and coronal mass ejection protons and up to a few GeV for solar flare protons.

The spectrum of electromagnetic radiation of the sun is also diverse. Using the experimental data for the solar spectral irradiance in wide range of wavelengths from 119.5 nm (~ 10.4 eV) up to more than 5000 nm (~ 0.25 eV), with the maximum around 500 nm from Ref. [8] we present the dependence of the spectral irradiance of solar electromagnetic radiation on the energy of solar photon shown in Fig. 1. (The total area under the curve in Fig. 1 gives the value of the solar constant $W_0$=1366.1 W/m$^2$). Almost 7% of the solar electromagnetic radiation consisting from ultraviolet (UV), $X$ - and $\gamma$ -rays [9] takes part only in an ionization of the sail producing the surface charge distribution. The visible wavelengths of the incident solar electromagnetic radiation are mostly reflected by the solar sail depending on the coefficient of reflection of the solar sail material. But a part of the visible radiation as well as the infrared portion of the spectrum with wavelengths are greater than 780 nm (~ 1.6 eV) will be mostly absorbed by the solar sail causing the heating of the sail and, therefore, the increase of its temperature. It is important to mention that about 47% of the solar radiation is in the range of infrared and microwave spectrum which is causing the heating of the sail and, as a result, its temperature increase.

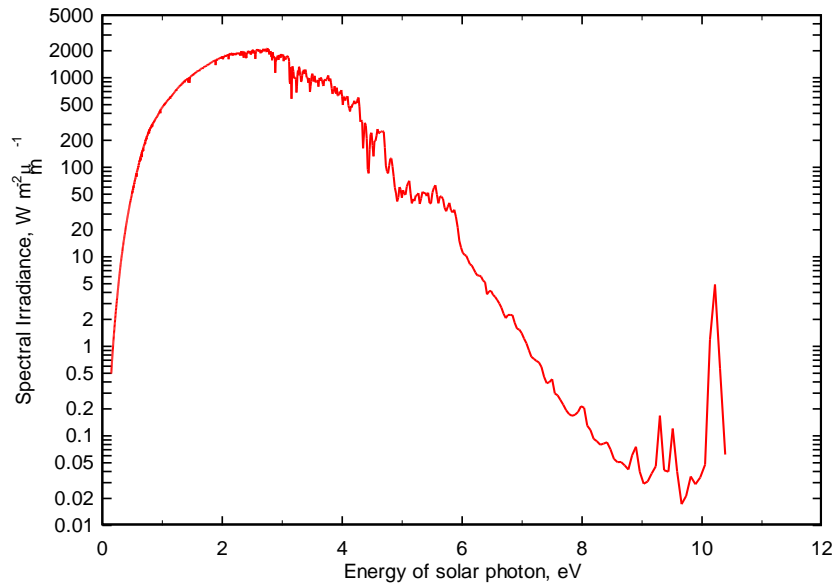

Fig. 1 Dependence of the Spectral Irradiance of solar electromagnetic radiation on the energy of a solar photon.

The interaction of the solar electromagnetic radiation with the solar sail not only accelerates the spacecraft but also induces the diversity of physical processes that change the physical properties of the sail. The interaction of the electromagnetic radiation with the solar sail material results in the photoelectric effect, the Compton effect and creation of electron-positron pairs that are the main processes that play a crucial role in the attenuation of the ultraviolet, X-ray and γ-ray by a solar sail. When the photons energy exceeds the threshold for nucleon knockout from solar sail material nuclei, a nuclear photoeffect occurs. All photons that produce the photoelectric effect or create the electron-positron pairs give up all their energy to the sail. For high-energy photons, which are $X$ - and $\gamma$ -rays portion of the spectrum, the sail is mostly transparent. However, metallic sails will be partially ionized by the solar UV radiation, as is shown in Ref. 10.

Consequently, we can conclude that basic components of the solar radiation that will interact with the solar sail are electromagnetic radiation with energy from a few tenths of eV to hundreds of MeV, and electrons and protons in the energy spectrum from a few tenths of eV to hundreds of MeV and up to GeV. Thus, the ionization of the sail, induced by the solar electrons, protons and helium ions, and its heating, resulted from the absorption of the electromagnetic waves, are two main culprits in degradation of the sail material that required a consideration.

Below the degradation of solar sail material is considered under the following assumptions:
- The only sources of degradation of the solar sail material are the solar electromagnetic radiation and corpuscular part of the solar radiation: electrons, protons, helium ions (α-particles). The cosmic radiation that mostly includes electrons and protons can be considered as a part of the solar corpuscular radiation.
- The optical parameters such as a reflection coefficient, an absorption coefficient and a solar sail's surfaces emission coefficients depend on the sail temperature. However, the optical coefficients do not depend on the incidence angle of electromagnetic radiation. Also, we are not considering the solar events. Therefore, the solar electromagnetic and corpuscular fluxes do not depend on time.

### III. Temperature Dependence of Solar Sail Parameters

**A. Thickness of a solar sail material**

The effect of ionizing radiations depends on both the radiation dose and the parameters of the solar sail material including the film thickness. It is already an established fact that the degradation is much more severe for the higher radiation doses and thinner films [11]. From the other hand one of the key design parameter, which determines the solar sail performance, is the solar sail areal mass, which depends on the thickness and the density of the sail material as follows:

$$s = \rho d \qquad (1)$$

where $d$ is the thickness of the sail and $\rho$ is the density of the sail material. An optimal control of solar sails depends on the sail areal mass as well as the sail pitch angle. It is clear from Eq. (1) that to obtain a high performance sail we should select among the materials with the same optical properties, the material with low density and use a thin film of this material for the solar sail. On the other hand, the degradation of a solar sail through the ionization also depends on the thickness of a sail material. It is obvious that when thickness of the sail increases the penetration of the solar radiation decreases and, therefore, the attenuation increases that causes the degradation of the solar film. If the solar sail film is too thin its degradation by electrons and protons decreases, however, it may become transparent for a part of the electromagnetic spectrum of the solar electromagnetic radiation and, therefore, this part of solar radiation is lost for the acceleration of the sailcraft. *The question is how thin the film for a solar sail should be that it would produce the acceleration of the sailcraft based on the maximum reflection of solar electromagnetic radiation and its optical degradation by UV radiation, X, γ – rays and the corpuscular solar radiation would be minimal?* For the electromagnetic part of the solar radiation electromagnetic fields inside a metallic conducting foil decay rapidly with depth. The distance it takes to reduce the amplitudes of the electromagnetic field by factor of $1/e$ (*e*-folding distance) is a skin depth and it is a measure of how far an electromagnetic wave penetrates into the conducting metallic foil. It is obvious that the foil thickness should be always larger than the skin depth; otherwise the solar sail material will be transparent to the electromagnetic radiation. Following the standard electrodynamics approach [12, 13] by applying the system of Maxwell's equations for linear conducting media in Refs. [14, 15] was found the minimum foil thickness that provides the maximum reflection and investigated dependence of this minimum thickness on wavelengths of solar electromagnetic radiation as well as on temperature. The condition for the minimum thickness of the solar sail that performs the acceleration of the sail based on the maximum reflection of the solar radiation should be at least the following [15]:

$$d = \omega^{-1}\left[\frac{\varepsilon(\omega)\mu_0}{2}\left(\sqrt{1+\left(\frac{\sigma(\omega,T)}{\varepsilon(\omega)\omega}\right)^2}-1\right)\right]^{-1/2}, \qquad (2)$$

where $\omega$ is a frequency of solar radiation, $\varepsilon(\omega)$ is the permittivity which is a function of the frequency and $\sigma(\omega,T)$ is the electrical conductivity of a material, which depends on the frequency (wavelength) of electromagnetic radiation and temperature. Using Eq. (2) we can estimate the minimal thickness of the solar sail required to achieve maximum reflection of the solar electromagnetic radiation for the given optical properties of the metallic foil. The required thickness of the metallic foil is governed by two properties of the solar sail material: the permittivity $\varepsilon(\omega)$ and the conductivity $\sigma(\omega,T)$ of the metallic foil. However, it is not the entire story. Actually, each of these parameters determines the optical properties of the solar sail material and depends to some extent on the frequency of the electromagnetic wave and the temperature. Indeed, Eq. (2) shows that the minimal thickness has the explicit dependence on the frequency through the factor $1/\sqrt{\omega}$ and it also has implicit dependence on frequency, because the conductivity $\sigma(\omega,T)$ and dielectric function $\varepsilon(\omega)$ are frequency dependent. Also the minimal thickness depends implicitly on the temperature through the temperature dependence of the conductivity.

The required thickness of the solar sail material is fluctuating depending on the temperature and frequency, therefore, we need to determine the minimal thickness of the sail material that would enable it to achieve the reflection of all solar spectrum frequencies at the temperatures that it got in the space environment. In studying the dependence of the sail material thickness on the temperature and on the solar electromagnetic spectrum frequency (wavelength) we use Eq. (2) and have assumed the solar sail material to be nonmagnetic, therefore, the permeability $\mu_0 = 4\pi \times 10^{-7}$ T·m/A.

The calculations performed using for the beryllium and aluminum films to find the required thicknesses of the solar sail material providing the best reflection of electromagnetic radiation as a function of wavelength of the solar radiation and temperature show that the general behavior of this dependence is such that the increase in the temperature requires increases in the thickness of the sail foil to keep its best reflection ability [15]. Also the analysis shows that the thickness of the foil exhibits the negligible dependence on the wavelength at low temperature and weakly decreases for all wavelength ranges at high temperature. However, much stronger dependence of the minimal thickness on the temperature, especially in the range corresponding to the visual part of the solar radiation spectrum. Both solar sail materials exhibit strong dependence of the thickness on the wavelength in the wavelengths range $0.2\,\mu\text{m} < \lambda < 0.8\,\mu\text{m}$ and this dependence becomes stronger when temperature increases. The analysis of the results of calculations shows that the minimal thickness requirement that provides the best reflection and absorption of all solar radiation wavelengths for temperature range up to about 800-900 K for which the limit of the elastic deformation can be reached, is about 50 nm for beryllium and aluminum that corresponds to the areal mass 0.1 g/m$^2$ and 0.14 g/m$^2$ for beryllium and aluminum, respectively.

Above we mentioned the results of calculation of the temperature dependence of the minimal required thickness for the solar sail film for the constant value for the temperature coefficient of conductivity over the range of temperature. However, in reality, the temperature coefficient of conductivity is not a constant over the above range of temperature. It increases with a rise of temperature but not linearly [21]. To demonstrate the importance of this fact, we used the experimental values of the temperature coefficient of conductivity from Ref.

[21] obtained for beryllium, extrapolated them, and made model calculations for the temperature dependence of the thickness of the beryllium film. The result of calculations shows that consideration of the temperature dependence of the temperature coefficient of conductivity leads to the increase of the required thickness by more than 35% at high temperatures [14]. Therefore, this factor must be also taken into consideration when designing the solar sail.

## B. Optical Parameters

The treatment of the optical properties of material must be properly done by quantum theory. However, for many purposes, an adequate and insightful treatment can be given in purely classical electron theory. There are four radiative properties of the material surface: spectral reflectivity $\rho(\lambda,T)$, absorptivity $\alpha(\lambda,T)$, transmissivity $\tau(\lambda,T)$ and emissivity $\varsigma(\lambda,T)$. All four properties are functions of temperature as well as of frequency of electromagnetic radiation. Absorptivity, reflectivity and emissivity can be evaluated in the framework of electromagnetic wave theory using the complex index of refraction, therefore, by using the electrical conductivity and the dielectric function of the material which must be known over the temperature and spectral range of interest. The electrical conductivity and dielectric function of the material are the functions of the wavelength of radiation and temperature. Such dependence for the case of the solar sail has been studied in detail in Refs. [14, 15]. Hence, the spectral absorptivity, reflectivity and emissivity of the surface of a solar sail should depend on the wavelength of radiation and temperature. However, the total absorptivity, reflectivity and emissivity of the surface are spectrally average value of the spectral absorptivity, reflectivity and emissivity and they depend on the temperature of the surface.

Let us consider the solar electromagnetic radiation traveling through the vacuum and hitting the sail surface of a conducting medium normally to the surface. Introducing the complex index of refraction as $n - im$ and taking into account that the solar electromagnetic radiation is unpolarized, the Fresnel's equation for the spectral reflectivity can be written in the following form [13]:

$$\rho_\lambda(\lambda,T) = \frac{[n(\lambda,T) - 1]^2 + m(\lambda,T)^2}{[n(\lambda,T) + 1]^2 + m(\lambda,T)^2}, \qquad (3)$$

where $n(\lambda,T)$ and $m(\lambda,T)$ are the indices of refraction and absorption, respectively and they depend on the wavelength and temperature and related to the real and imaginary parts of the complex wave number given in Ref.[14] as

$$n(\lambda,T) = \frac{ck}{\omega} = \sqrt{\frac{\varepsilon(\omega)\mu}{2\varepsilon_0\mu_0}} \left[ \sqrt{1 + \left(\frac{\sigma(\omega,T)}{\varepsilon(\omega)\omega}\right)^2} + 1 \right]^{1/2}, \quad m(\lambda,T) = \frac{c\kappa}{\omega} = \sqrt{\frac{\varepsilon(\omega)\mu}{2\varepsilon_0\mu_0}} \left[ \sqrt{1 + \left(\frac{\sigma(\omega,T)}{\varepsilon(\omega)\omega}\right)^2} - 1 \right]^{1/2}. \quad (4)$$

For opaque media from Kirchhoff's law that states an object must emit at the same rate as it absorbs if equilibrium is to be maintained, follows that the spectral emissivity related to the spectral reflectivity as $\varsigma_\lambda(\lambda,T) = 1 - \rho_\lambda(\lambda,T)$. Therefore, from (3) it is easy to obtain that

$$\zeta_\lambda(\lambda,T) = \frac{4n(\lambda,T)}{[n(\lambda,T) + 1]^2 + m(\lambda,T)^2}. \qquad (5)$$

For description of the optical properties of metals it is convenient to introduce the complex dielectric function

$$\varepsilon(\omega) = \varepsilon - \frac{i\sigma(\omega,T)}{\omega} \qquad (6)$$

This function is related to the real and imaginary parts of the complex index of reflection in the following way

$$[n(\lambda,T) - im(\lambda,T)]^2 = \varepsilon(\omega) - \frac{i\sigma(\omega,T)}{\omega}, \qquad (7)$$

Therefore,

$$n(\lambda,T)^2 - m(\lambda,T)^2 = \varepsilon(\omega), \quad n(\lambda,T)m(\lambda,T) = \frac{\sigma(\omega,T)}{2\omega}. \qquad (8)$$

From (8) we can express $n(\lambda,T)^2$ and $m(\lambda,T)^2$ through the measurable electrical conductivity as

$$n(\lambda,T)^2 = \frac{1}{2}\left(\varepsilon(\omega) + \sqrt{\varepsilon(\omega)^2 + \left(\frac{\sigma(\omega,T)}{2\omega}\right)^2}\right), \quad m(\lambda,T)^2 = \frac{1}{2}\left(-\varepsilon(\omega) + \sqrt{\varepsilon(\omega)^2 + \left(\frac{\sigma(\omega,T)}{2\omega}\right)^2}\right). \qquad (9)$$

The value of $n(\lambda,T)^2$ and $m(\lambda,T)^2$ can be predicted for all wavelengths and temperature or the corresponding values of $\varepsilon(\omega)$ and $\sigma(\omega,T)$ can be determined within certain theoretical model or measured experimentally.

Let us now consider the frequency dependence of the conductivity. The best way to do that is to use the classical Drude model for conductivity [16]. According to the Drude model, the frequency dependent conductivity is a complex function and given by

$$\sigma(\omega) = \frac{\sigma(T)}{1-i\omega t}, \quad (10)$$

where $\sigma(T)$ is the DC Drude conductivity at temperature $T$ and the relaxation time $t$ for metal is about $t = 10^{-14}$ s. In Drude model we can consider

$$n^2 = m^2 \approx \frac{\sigma(T)}{\varepsilon_0 \omega} \quad (11)$$

and (3) and (5) become

$$\rho_\lambda(\lambda,T) = \frac{2n(\lambda,T)^2 - 2n(\lambda,T) + 1}{2n(\lambda,T)^2 + 2n(\lambda,T) + 1}, \quad (12)$$

$$\varsigma_\lambda(\lambda,T) = \frac{4n(\lambda,T)}{2n(\lambda,T)^2 + 2n(\lambda,T) + 1}. \quad (13)$$

The electrical conductivity depends on temperature and *approximately inversely* proportional to it: $\sigma(T) = \sigma_0 / [1+\alpha(T-T_0)]$, where $\sigma_0$ is DC conductivity at temperature $T_0$ and $\alpha$ is the temperature coefficient of the conductivity. Hence, $n(\lambda,T)^2$ also depends on temperature, and therefore, spectral reflectivity and emissivity will show temperature dependence, however in different ways. Because $\rho_\lambda(\lambda,T)$ is the ratio of the same order polynomials the dependence on $T$ in the denominator and the numerator in (12) cancels each other and it becomes negligible, but we cannot say the same about the emissivity. As it is seen from (13) the emissivity $\varsigma_\lambda(\lambda,T)$ is approximately directly proportional to $T^{1/2}$. It is easy to show that from the following expansion. Since $n>1$ (12) and (13) may expanded as

$$\rho_\lambda(\lambda,T) = \frac{2n(\lambda,T)^2 - 2n(\lambda,T) + 1}{2n(\lambda,T)^2 + 2n(\lambda,T) + 1} \approx 1 - \frac{2}{n(\lambda,T)} + \frac{2}{n(\lambda,T)^2} + \ldots =$$
$$1 - 2\sqrt{\frac{\varepsilon_0 \omega [1+\alpha(T-T_0)]}{\sigma_0}} + 2\frac{\varepsilon_0 \omega [1+\alpha(T-T_0)]}{\sigma_0} + \ldots \quad (14)$$

$$\varsigma_\lambda(\lambda,T) = \frac{4n(\lambda,T)}{2n(\lambda,T)^2 + 2n(\lambda,T) + 1} \approx \frac{2}{n(\lambda,T)} - \frac{2}{n(\lambda,T)^2} + \ldots =$$
$$2\sqrt{\frac{\varepsilon_0 \omega [1+\alpha(T-T_0)]}{\sigma_0}} + 2\frac{\varepsilon_0 \omega [1+\alpha(T-T_0)]}{\sigma_0} + \ldots \quad (15)$$

The consideration of the leading terms in (14) and (15) demonstrates that the spectral reflectivity weakly depends on temperature, emissivity depends on temperature and increases when temperature increases. The total hemispherical emissivity is can be approximated by following equation [17]

$$\varsigma(T) = 7.66 \left(\frac{T}{\sigma(T)}\right)^{1/2}. \quad (16)$$

For Al the coefficient is 7.52 K$^{-1/2}\Omega^{-1/2}$m$^{-1/2}$ [18]. Since the electrical conductivity $\sigma$ is approximately inversely proportional to the temperature, therefore, the hemispherical emissivity, as well as the total normal emissivity is approximately linearly proportional to the temperature. Eq. (16) satisfactorily describes the temperature dependence of the experimental data for the hemispherical emissivity for various metals [17].

In general, the solar sail is the sandwich of films and front and back sides of the solar sail will have different surface emissivity. According to the Stefan-Boltzman's law the rate of energy emitted from a unit area of the surface material at temperature $T$ is proportional to the fourth power of the temperature of the surface. The solar sail radiates heat into the space that has a temperature 2.725 K related to the cosmic micrrowave background radiation. For simplicity we consider that this temperature is 0 K. Taking this fact into consideration for the rate of energy emitted by both sides of a unit area of the solar sail we obtain

$$W_e = [\varsigma_f(T) + \varsigma_b(T)]\sigma_{SB} T^4, \quad (17)$$

where $\varsigma_f(T)$ and $\varsigma_b(T)$ are the front and back sides surface emissivity of the sail at given temperature, respectively and $\sigma_{SB} = 5.67 \times 10^{-8}$ Wm$^2$K$^{-4}$ is Stefan-Boltzman constant. Based on the law of conservation of energy the solar sail will be in thermal equilibrium if the total absorbed energy equals to total emitted energy. This condition allows finding the thermal equilibrium temperature [19, 20]

$$T = \left( \frac{[1-\rho(T)-\tau(T)]W_0 R_0^2 \cos\theta}{[\varsigma_f(T)+\varsigma_b(T)]\sigma_{SB}} \right)^{1/4} r^{-1/2}. \quad (18)$$

Eq. (18) is obtained under the assumption that solar energy flux has an inverse square variation with the distance from the Sun (in fact, this assumption is not valid when account is taken of the finite angular size of the solar sail [20] and considering that the solar energy flux with the angle of an incidence $\theta$ to the surface of the sail at the distance $r$ from the Sun) and the solar irradiance $W_0$ at $R_0 = 1$ AU. From Eq. (18) it follows that the temperature of the solar sail material is increases when the solar sail approaches to the sun as $T \sim r^{-1/2}$. Taking into account the temperature dependence of the emissivity (17) (emissivity is approximately linearly proportional to temperature) in Eq. (18) we can conclude that the temperature increases approximately as $T \sim r^{-2/5}$, when the heliocentric distance $r$ decreases. In other words, when we are taking into account the temperature dependence of the emissivity (17) and a conductivity of metals the temperature dependence on the distance is not the same as for the constant emissivity and conductivity. In this case the temperature of the solar sail material increases more slowly when sail approaches to the sun than in the case for the constant emissivity and conductivity. However, the increase of the temperature decreases the reflectivity of the solar sail which is a disadvantage for its dynamics (acceleration). Can the disadvantage related to the temperature be turned into the advantage for the solar sail dynamics? The hot sail may have one advantage related to the thermal desorption propulsion method proposed in Ref. [21]. The *thermal desorption* is the process for mass loss and dominates all other processes for mass loss above temperatures of 300–500 $^0$C. If we can heat sails to temperatures >1000K and the surface of solar sail has a coat of embedded atoms or paint, their *thermal desorption* can provide higher specific impulse than liquid rockets. In other words, atoms embedded in a substrate can be liberated by hear heating and provide an additional trust. Using the thermal desorption for thrust is not a new idea, but it is new to apply this idea to the solar sail that naturally gains temperature through the absorption of solar radiation. Heating of a sail by solar radiation to the temperature until its surface coat sublimes or desorbs and, therefore, can add far more thrust sounds attractive. However, it requires the search and surveys of promising new materials for thermal desorption that can be embedded on the surface of solar sail, study their optical properties and usage for thrusting applications.

## IV. Degradation of Solar Sail Materials by Solar Radiations

The solar sail is exposed to different environmental factors that cause the degradation of the solar sail materials which can be a composite multilayer system of high reflective films on the polymeric substrates. The main factors of space environment that lead to the degradation of the solar sail material are vacuum UV radiation flux of electrons, protons and α-particles radiated by the sun.

### A. UV Radiation

Vacuum ultraviolet radiation is one of the major environment factors causing the degradation of solar sail material. Although the UV energy percentage in the total spectrum of the Sun is very limited, its photon energy is high enough to significantly change reflectivity of the metallic films and break most chemical bonds in the polymers [23]. Under UV radiation, the mechanical properties as tensile properties of metallic film did not change significantly [24, 25], while the optical ones varied noticeably. With increasing the irradiation dose, the tensile fracture strength and elongation decreased slightly, and the spectral absorbance increased remarkably in the ultraviolet to visible regions.

The intensity of UV, $X$-ray, and $\gamma$-ray radiations penetrating a layer of a solar sail of material with thickness $d$ is given by the exponential attenuation law $I = I_0 e^{-\mu\rho d}$, where $I_0$ is the incident photons intensity and $\mu$ is a mass attenuation coefficient. Using Eq. (1) the last equation can be rewritten in the following form:

$$I = I_0 e^{-\mu s}. \quad (19)$$

Eq. (19) gives the relationship between the key parameter $s$ that **determines the performance** of the solar sail and the mass attenuation coefficient $\mu$ of the photon flux that related to the degradation of the surface of the solar sail. The mass attenuation coefficient $\mu$ is a basic quantity used in calculations of the penetration and the energy deposition by photons in materials that leads to degradation of optical properties of the solar sail. The mass attenuation coefficient is directly related to the total cross section of photon interaction with atoms of the solar sail material. The total cross section can be written as the sum of contributions from the principal photon interactions with a metal film: $\sigma_{ape}$ is the atomic photoelectric effect cross section, $\sigma_C$ is the Compton scattering cross section, $\sigma_{pair}$ is the cross section for electron-positron production, and $\sigma_{npe}$ is the photonuclear cross section. As a result for the mass attenuation coefficient we have

$$\mu = (n\sigma_{ape} + nZ\sigma_C + n\sigma_{pair} + n\sigma_{npe}), \qquad (20)$$

where $n$ is the number of metal atoms in the unit volume. The attenuation coefficient is a function of photon energy because each cross section depends on the photon energy. The first term in Eq. (20) is dominant for a low photon energy, the second term becomes important for photon energy up to hundreds of keV, the third term plays the major role for photons' energy higher than a few MeV, and the last term gives the dominating contribution for high-energy photons when they are interacted with nuclei of the solar sail material. Using the corresponding cross section for the considered processes we can estimate the reduction of the radiation through the solar sail. The results of the attenuation of the photons by beryllium and aluminum thin foils are presented in Fig. 2. It is important to mention that for the same areal mass of the solar sail the thickness of the aluminum film is 1.46 times larger than that for the beryllium film that is equal to the ratio of the densities of aluminum and beryllium. To make the most conservative prediction of the dependence of the ratio $I/I_0$ on the areal mass of the beryllium and aluminum films for the ionization processes, we use the maximum value of the attenuation coefficients for the corresponding processes. The results presented in Fig. 2 indicate that UV radiation with energy less that 10 eV are completely absorbed by Al and Be films when the areal mass more than 0.5 g/cm$^2$. It is interesting to mention that for the photon energy range between about 15 eV and 100 eV the ionization of aluminum is always less than that for beryllium. This is due to the fact that mass attenuation coefficient for Al at these photons energies is less than one for the beryllium. For photon energies more than 1keV ionization of the beryllium is relatively smaller than of the aluminum. However, at these energies less than 50% of photons penetrate the sail foil for both metals if the areal mass equals 5 g/m$^2$. At the energy of photons below 150 eV the major effect of the attenuation of the photons is the ionization of the metallic surface of the solar sail. Therefore, the solar sail is gaining the positive electric charge. Also our estimate based on the values of the mass attenuation coefficient at high energies shows that less than 1% of 1 MeV photons attenuated by each of films with the areal mass more than 1 g/cm$^2$. When the film areal mass increases the degree of ionization increases also as it was demonstrated in Ref. 26 for different candidates solar sail material.

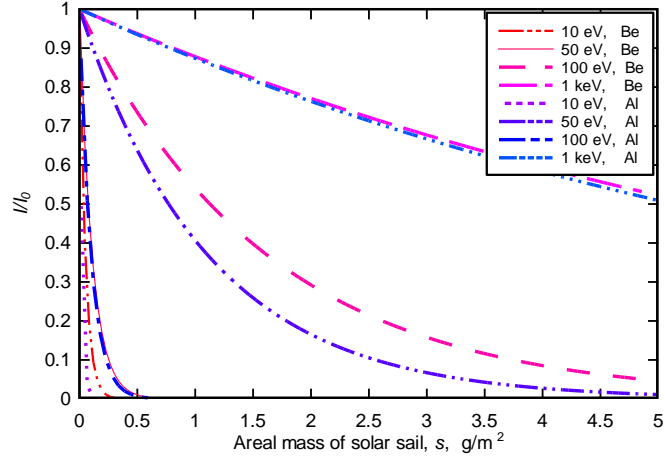

Fig. 2  Attenuation of photons by beryllium and aluminum films for different photon energies.

**B. Degradation by solar electrons, protons and α-particles**

Electrons, protons and α-particles will also interact with the solar sail film. At low energy, electrons scatter on the sail material atoms and basic physical processes are the excitations and ionization of the atoms of solar sail material. We can use Eq. (19) to determine the solar electron flux reduction; however, the mass attenuation coefficient for electrons is defined as

$$\mu_e = (n\sigma_{ex} + nZ\sigma_i), \qquad (21)$$

where $\sigma_{ex}$ is the sum of total cross sections for excitation processes in different states, and $\sigma_i$ is the sum of total cross sections for the ionization with excitation. The energy dependence of the cross section results in the energy dependence of the mass attenuation coefficient for the electron. The scenario with the maximum value of the cross section for each process is conservatively assumed to better understand the processes' influence on the ionization of the solar sail surface. To make a meaningful estimate of all effects of the interaction of photons with the solar sail material leading to its ionization, we consider the upper limits of the experimental and theoretical values of the cross sections for all considered processes.

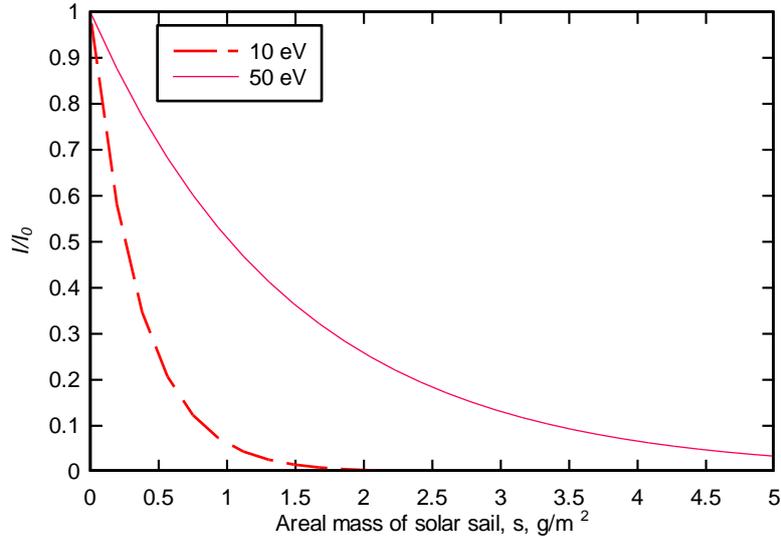

Fig. 3 The dependence of the ratio $I/I_0$ on the beryllium areal mass for different electron energies.

As an example we consider the beryllium and calculate the dependence of the ratio $I/I_0$ on the areal mass for different electron energies shown in Fig. 3. The results show that 10 eV electrons are stopped by the beryllium film with the areal mass 1.5 g/cm$^2$, while about 50% of 50 eV electrons penetrate the beryllium film with this areal mass. The stopped electrons mostly ionize the beryllium film and lead to the excitation of beryllium atoms that finally results their ionization through the different auto ionization channels.

When electrons, protons and α-particles pass through the sail material there is the energy loss for electrons, protons and α–particles. For electrons, this energy loss is the result of collisions with background atomic electrons and bremsstrahlung radiation when they are strongly deflected by the nuclei of solar-sail material atoms. For protons and α–particles, it is caused by electronic and nuclear collisions. Figs. 4 and 5 present the dependence of the stopping range for protons and α-particles on the energy of proton and α-particle. We use the data from the ESTAR and ASTAR database [27] and the program, which calculates the stopping range for protons and α-particles in beryllium and aluminum. The comparison of Fig. 4 and 5 shows that the stopping range of the proton in beryllium is larger than that in aluminum, while there the situation is opposite for α-particle. The stopping range for α-particle in aluminum always exceeds that for the beryllium. From the presented result we can conclude that protons and α-particles with energies from 10 keV to 100 keV 1 MeV will be stopped by the beryllium or aluminum solar sail foil with the areal mass between 0.1 g/m$^2$ and 10 g/m$^2$. This results in ionization of the solar sail film and degradation its optical properties. Let us also mentioned that the stopping range much lower for the beryllium than for aluminum, scandium, titanium and niobium, which also are considered as candidate solar-sail materials [28].

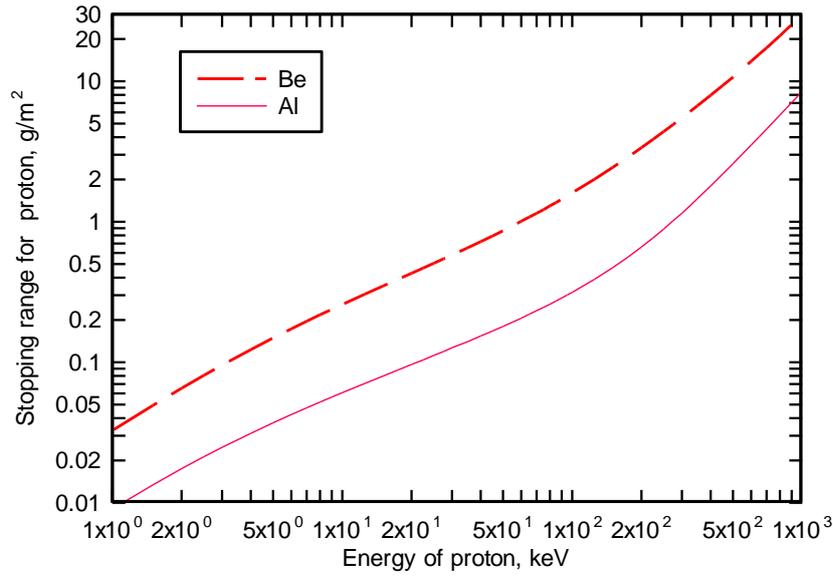

Fig. 4 The dependence of the stopping range of proton on the energy.

The interaction of photons, electrons, protons and α-particles with solar sail material produces atomic displacements, electronic excitations, or both. Atomic displacements result from the elastic scattering of the protons and α-particles by an atomic nucleus so that the kinetic energy transferred to the nucleus in the collision is sufficient to break the chemical bonds to neighboring atoms. Inelastic scatterings of these particles produce macroscopic modifications in the properties of solids that are dissimilar to those caused by the electrons. Electron-induced displacement damage in materials is qualitatively and quantitatively unlike that caused by the protons or α-particles.

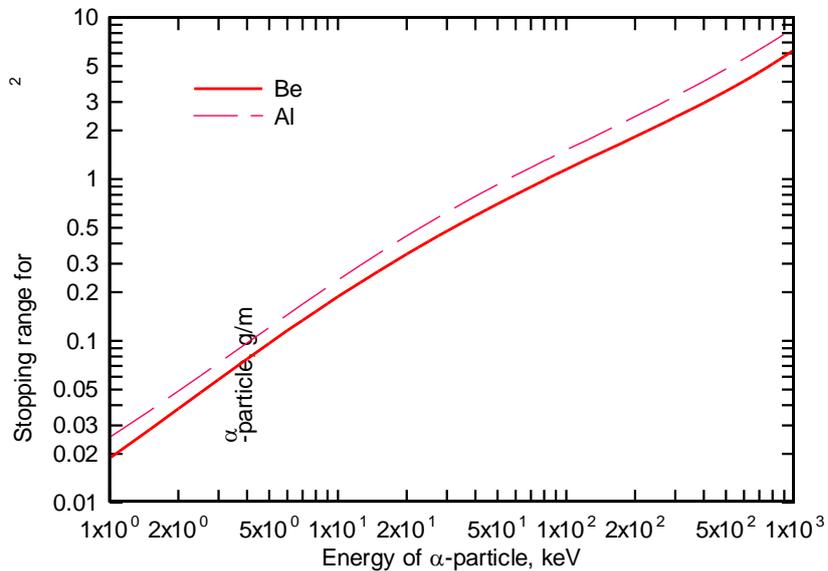

Fig. 5 The dependence of the stopping range of α−particle on the energy.

## V.  Discussion and Conclusions

Functional solar sails must be resistant to the effects of space environmental exposure. The influence of the effects of space environmental factors such as a vacuum UV radiation, electrons, protons and helium ions impact on materials were studied experimentally [1, 4-6, 29-30]. These experimental studies focused on the measurements of the hemispherical reflectance when samples under investigation have been exposed to electron, proton, and UV irradiation separately and combined were investigated. Peculiarities in changes of spectral reflectance of thermal control coatings under single and multiple influence of space environment were studied. Exposure of samples to environmental factors, separately and in combination, demonstrated different dependence of the spectral reflectance coefficient on the exposure that is associated with various processes of defect generation. The most significant irreversible changes of optical properties of materials are produced by protons. In this case the reflectance of materials changes appreciably. The most important conclusion that space environmental factors demonstrate the effect of nonadditivity: the influences of separate environmental factors and their combined action are not the same. One should take this fact into account in theoretical modeling of the space environmental effects.

We have discussed the influence the of space environment on the solar sail temperature that it gains as a result of absorption of solar radiation. The increase of temperature slightly change the reflectivity of the metals used in aerospace industry, and not significantly affect the propulsion efficiency of the sail. However it requires having more thick material to reflect all wave lengths of electromagnetic radiation at high temperature than when solar sail is at low temperature. Of course this increases the areal mass affects significantly on the performance of solar sail. Therefore, the strategy of solar sail mission have to consider this fact up prior based on a projected maximum temperature that solar sail will reach in space depending on the distance from the sun. Results of our study show that the temperature of a solar sail increases approximately as $T \sim r^{-2/5}$, when the heliocentric distance $r$ decreases, when are taking into account the temperature dependence of the emissivity and conductivity of the solar sail material. Thus, the temperature of the solar sail material increases more slowly than in the case for the constant emissivity and conductivity ($T \sim r^{-1/2}$).

In the present work is demonstrated that the space environment related to UV radiation, electrons, protons and helium ions influence:
1. Structural changes in the surface of solar sail that results in the radiation-enhanced adsorption, decrease of reflection and desorption. The radiation-enhanced adsorption leads to the increase the solar sail temperature.
2. Generation of the electric field due to accumulation of the volume charge at the materials surface under the action of ionizing radiation. The effect of ionizing radiations depends on the radiation dose.

Thus, the temperature increase and the accumulation of the positive electric charge are the most two disadvantages that causes by the space environmental factors. Let us address the issue how we can turn these disadvantages into the advantages. The ionized charged sail may have advantage related to the electrical propulsion by the solar wind proposed in Ref. [31]. As is shown above in our case as a result of ionization the solar sail naturally gains the positive charge (potential) which will deflect the proton component of the solar wind and therefore, extract momentum from the solar wind plasma. In Refs. [26, 32] was analyzed such opportunity for the beryllium hollow body solar sail. However, the proton component of the solar wind induced acceleration is considerably less that the solar radiation-pressure acceleration. When solar sail naturally gains the heat its temperature increases and can be so high that can be used the thermal desorption propulsion method proposed in Ref. [21]. Using coats of atoms, that liberate by hear heating at high temperature through thermal desorption, embedded in the reflecting surface of solar sail can provide an additional significant trust for a solar sail. After the coats sublime away, the sail can perform as a conventional solar sail, using a beryllium or aluminum coat beneath.

The determination of optical properties of the solar materials remains a challenging task, despite the advanced methods of data acquisition and analysis. It is required to develop mathematical model that allows model the time dependence of optical parameters of the solar sail on dose of radiation exposed by the sun and use these parameter in orbital equation for a realistic mission analysis. In Ref. [33] are made some steps toward mathematical models describing the influence of electron radiation on outgassing of spacecraft materials. The theoretical works has been accomplished in the aria of an imperfectly reflecting solar sail dynamics modeling and considering optical degradations of the solar sail [2, 34, 35]. However, it was considered the static model assuming that the optical parameters are changed as a result of the space environment but not undergo the changes during the mission. Unique technology involving optical performance, environmental stability, thermal expansion and wrinkle management of the sail film, system dynamics and control strategies and many other issues must be addressed for realistic minimum mass solution [36]. This is problematic but a challenging task.

## References

[1] Edwards, D. L., Semmel, C., Hovater, M., Nehls, M., Gray, P., Hubbs, W., And Wertz, G., "Status Of Solar Sail Material Characterization At Nasa's Marshall Space Flight Center," *Protection of Materials and Structures from Space Environment,* edited by J.I. Kleiman, Springer, 2006, pp. 233–246.
[2] Dachwald, B., and Macdonald, M., Parametric Model and Optimal Control of Solar Sails with Optical Degradation, Journal of Guidance, Control, and Dynamics Vol. 29, No. 5, 2006, pp. 1170–1178.

doi:10.2514/1.20313.

[3] Nuclear and Space Radiation Effects on Materials, NASA Space Vehicle Design Criteria, NASA SP-8053, 1970.

[4] Edwards,D., Hubbs,W., Gray, P.Wertz, G., Hoppe,D., Nehls, M., Semmel, C., Albarado, T., and Hollerman,W., In *Proceedings of the 9th International Symposium on Material in a Space Environment*, Noordwijk, The Netherlands, ESA Publications, Noordwijk, The Netherlands June 2003, pp. 16–20.

[5] Albarado, T., Hollerman, W., Edwards, D., Hubbs, W., and Semmel, C. (2003) In *Proceedings of ISEC 2003: 2003 International Solar Energy Conference*, Hawaii, 15–18 March 2003.

[6] Edwards, D., Hubbs, W., Stanaland, T., Hollerman, A., and Altstatt, R. (2002) In *Proceedings of SPIE Photonics for Space Environments VIII*, Vol. 4823, 2002.

[7] Dalla Vedova, F., Henrion, H., Leipold, M., et. al. "The Solar Sail Materials (SSM) project – Status of activities", Advances in Space Research Vol. 48, 2011, pp. 1922–1926.

[8] 2000 ASTM Standard Extraterrestrial Spectrum Reference E-490-00. [online database], URL: http://rredc.nrel.gov/solar/spectra/am0/. (Date Accessed 19 September 2008).

[9] Michalsky, J.J., "The Astronomical Almanac's algorithm for approximate solar position (1950–2050)", Solar Energy Vol. 40, 1998, pp. 227–235.

[10] Kezerashvili, R. Ya., and Matloff, G.L., "Solar radiation and the beryllium hollow-body sail: 1. The ionization and disintegration effects", JBIS, Vol. **60**, 2007, pp.169-179.

[11] Arshak, K., and Korostynska, O., *Mater. Sci. Eng.* Vol. B133, 2006, pp. 1-.

[12] H.C. Obanian, "*Classical Electrodynamics*", 2$^{nd}$ ed., Infinity Science Press LLC, 2007.

[13] D.J. Griffiths, "*Introduction to Electrodynamics*", 3$^{rd}$ ed., Prentice Hall, 1999.

[14] Kezerashvili, R.Ya., "Solar sail interstellar travel: 1. Thickness of solar sail films", JBIS Vol. 61, 2008, pp. 430–439.

[15] Kezerashvili, R. Ya., "Thickness requirement for solar sail foils", Acta Astronautica Vol. 65, 2009, pp. 507–518.

[16] Ashcroft, N.W., and Mermin, N. D., Solid State Physics, Brooks/Cole Thomson Learning, 1976.

[17] Parker, W.J., and Abbott, G.L., Theoretical and Experimental Studies of the Total Emittance of Metals, in Symposium on Thermal Radiation of Solids, ed. S. Katzoff, NASA SP-55, 1965, pp. 11-28.

[18] Latyev, L.N., Petrov, V. A., Čechovskoj, V. J., Šestakov, E. N. Izlutchatelnyje svojstva tverdych materialov. Moskva: Energia, (In Russian) 1974, pp. 306-310; pp. 400-409.

[19] Polyakhova, E.N., Kosmicheskii Polet s Solnechnim Parusom (in Russian) (Cosmic flight with solar sail), Nauka, Moscow, 1988.

[20] McInnes, C.R., Solar Sailing. Technology, Dynamics and Mission Applications, Springer, Praxis Publishing.1998.

[21] Benford, J., and Benford, J., Acceleration of sails by thermal desorption of coatings, Acta Astronautica Vol. 56, 2005, pp. 593 – 599.
doi:10.1016/j.actaastro.2004.09.049

[22] Lewis, E.J., "Some Thermal and Electrical Properties of Beryllium", Phys. Rev. Vol. **34**, 1929, pp.1575 – 1587.

[23] Peng, G., Yang, D., and He, S., Effect of VUV Radiation on Properties and Chemical Structure of Polyethylene Terephthalate Film, *Protection of Materials and Structures from Space Environment,* edited by J.I. Kleiman, Springer, 2006, pp. 225–232.

[24] Zhou, O., Tanaka, K., and Itoh, G., Deformation of aluminum thin foils under tensile stress at elevated temperature, J. Mater. Sci. Lett. Vol. **21,** 2002, pp. 215-216.

[25] Yu, D. Y. W., and Spaepen, F., The yield strength of thin copper films on Kapton, J. Appl. Phys. Vol. **95**, 2004 pp. 2991- 2997.

[26] Matloff, G.L., Kezerashvili, R.Ya. Interstellar solar sailing: a figure of merit for monolayer sails. JBIS Vol. 61, 2008, pp. 330–333.

[27] Berger, M.J., Coursey, J.S., Zucker, M.A., Chang, J., ESTAR, PSTAR, and ASTAR: Computer Programs for Calculating Stopping-Power and Range Tables for Electrons, Protons, and Helium Ions. NIST, Gaithersburg, MD, 2005.

[28] Kezerashvili, R. Ya., and Matloff, G.L., "Microscopic approach to analyze solar-sail space-environment effects", Advances in Space Research Vol. 44, 2009, pp. 859–869.

[29] Prosvirikov, V. M., A. V. Grigorevskiy, A.V., Kiseleva, L. V., Zelenkevich, A. P., And Tsvelev, V. M., "Influence of Space Environment on Spectral Optical Properties of Thermal Control Coatings" , *Protection of Materials and Structures from Space Environment,* edited by J.I. Kleiman, Springer, 2006, pp. 61–69.

[30]Khassanchine, R. H., A. N. Timofeev, A.N., Galygin, A.N., Kostiuk, V. I., And Tsvelev, V. M., "Influence of Electron Radiation on Outgassing of Spacecraft Materials", " , *Protection of Materials and Structures from Space Environment,* edited by J.I. Kleiman, Springer, 2006, pp. 43–50.

[31] Janhunen, P., Sandroos, "Simulation study of solar wind push on a charged wire: basis of solar wind electric sail propulsion", Ann. Geophys. Vol. 25, 2007, pp. 755–767.

[32] Kezerashvili, R. Ya., and Matloff, G.L., "Solar radiation and the beryllium hollow-body sail: 2. Diffusion, recombination and erosion processes". JBIS Vol. 61, 2008, pp. 47–57.

[33] Khassanchine, R. H., Grigorevskiy, A. V., and Galygin, A. N., "Influence of Electron Radiation on Outgassing of Spacecraft Materials", Journal of Spacecraft and Rockets Vol. 41(3), 2004, pp. 384–388.

[34] Rios-Reyes, L., and Scheeres, D. L., "Generalized Model for Solar Sails," Journal of Spacecraft and Rockets, Vol. 42, No. 1, 2005, pp. 182–185.

[35] Mengali, G., and Quarta, A. A., "Optimal Three-Dimensional Interplanetary Rendezvous Using Nonideal Solar Sail," Journal of Guidance, Control, and Dynamics, Vol. 28, No. 1, 2005, pp. 173–177.

[36] Murphy, D.M., Murphey, T.W., and Gierow, P.A., "Scalable Solar-Sail Subsystem Design Concept", Journal of Spacecraft and Rockets, Vol. 40, No. 4 (2003), pp. 539-547.